\documentclass{amsproc}

\newtheorem{theorem}{Theorem}[section]

\usepackage{graphicx}

\theoremstyle{definition}

\theoremstyle{remark}

\numberwithin{equation}{section}

\newcommand{\be}{\begin{equation}}
\newcommand{\ee}{\end{equation}}
\newcommand{\bea}{\begin{eqnarray}}
\newcommand{\eea}{\end{eqnarray}}
\newcommand{\eu}{{\rm e}}
\newcommand{\ii}{{\rm i}}
\newcommand{\de}{{\displaystyle\rm\mathstrut d}}
\newcommand{\Ord}{{\rm O}}

\newcommand{\Tr}{{\rm Tr}}

\begin{document}

\title{Entanglement Spectrum for the  $XY$ Model in One Dimension}

\author{F. Franchini}
\address{The Abdus Salam ICTP, Strada Costiera 11, Trieste (TS), 34014, Italy}
\curraddr{SISSA, Via Beirut 2-4, 34151 Trieste (TS), Italy}
\email{fabio.franchini@sissa.it}
\thanks{Fabio Franchini was supported in part by PRIN Grant 2007JHLPEZ.}

\author{A. R. Its}
\address{Department of Mathematical Sciences, Indiana
University-Purdue University Indianapolis, Indianapolis, IN
46202-3216, USA}
\email{itsa@math.iupui.edu}
\thanks{Alexander Its was supported by NSF Grant DMS-0701768}

\author{V. E. Korepin}
\address{C.N.\ Yang Institute for Theoretical Physics, State
University of New York at Stony Brook, Stony Brook, NY 11794-3840,
USA}
\email{korepin@insti.physics.sunysb.edu}
\thanks{Vladimir Korepin was supported by NSF Grant DMS 0905744}

\author{L. A. Takhtajan}
\address{Mathematics Department, State
University of New York at Stony Brook, Stony Brook, NY 11794-3840,
USA}
\email{leontak@math.sunysb.edu}
\thanks{Leon Takhtajan was supported in part by NSF Grant DMS-0705263}

\date{\today}


\begin{abstract}
We consider the reduced density matrix of a large block of consecutive spins in the ground states of the $XY$ spin chain on an infinite lattice.
We derive the spectrum of the density matrix using the expression of the R\'enyi entropy in terms of modular functions.
The eigenvalues $\lambda_{n}$ form an exact geometric sequence. For example, for strong magnetic field $\lambda_{n}= C \exp{(-\pi \tau_{0}n)}$, here $\tau_{0}>0$ and $C > 0$ depend on the anisotropy and the magnetic field.
Different eigenvalues are degenerated differently. The largest eigenvalue is unique, but the degeneracy $g_{n}$ increases sub-exponentially as eigenvalues diminish: $g_{n}\sim \exp{(\pi \sqrt{n/3})}$. For weak magnetic field expressions are similar.

\end{abstract}

\maketitle

\section{Introduction}

Entanglement is a peculiar feature of a quantum system, which  distinguishes it from a classical one. While the notion of entanglement has been introduced since the dawn of quantum mechanics, only recently physicists have employed a quantitative approach, mostly inspired by the progresses in  quantum information and Bethe ansatz.

The most studied quantity  is  the {\it Von Neumann entropy}, which is the quantum analog of the  Shannon entropy and measures the amount of (quantum) information stored in a system $A$ described by a density matrix $\rho_A$:
\be
   S(\rho_A) \equiv -\Tr (\rho_A \ln\rho_A) \; .
   \label{Sdef}
\ee
We shall consider the simplest case of a pure system $U$ described by a wave function $| \Psi \rangle$. The system $U$ is composed by the union of two disjoint subsystems $A$ and $B$, meaning $U \equiv A \cup B$.
The reduced density matrix of subsystem $A$ is obtained by tracing out the $B$ degrees of freedom $\rho_A \equiv \Tr_B | \Psi \rangle \langle \Psi |$ and the Von Neumann entropy of a subsystem is a measure of the entanglement between the two subsystems \footnote{note that  $S(\rho_A)=S(\rho_B)$ } \cite{NielsenChuang}.

Characterizing the entanglement with a single number is definitely appealing, but it can hardly capture its complexity. For this reason, other entanglement  measures have been introduced, such as,  the {\it Renyi entropy} \cite{renyi, abe, BD, hb, L}:
\be
   S_R (\rho_A, \alpha) \equiv \frac{1}{1-\alpha} \ln \Tr(\rho_A^{\alpha}),
 \qquad   \qquad  ~~~~1> \alpha > 0 \; .
   \label{Salphadef}
\ee
Note that in the limit $\alpha \to 1$, the Renyi entropy recovers the Von Neumann entropy: $$\lim_{\alpha \to 1} S_R (\rho_A, \alpha) = S (\rho_A).$$
We calculated the {\it Renyi entropy} of a large block of spins for the $XY$ spin chain in \cite{jin, renyipaper} and  studied its  analytical continuation  into the  complex plane of the parameter $\alpha$.

Another quantity that recently has attracted a lot of interest is the spectrum of the reduced density matrix, which is now referred to in the literature as the {\it entanglement spectrum}, after \cite{LiHaldane06}.
The knowledge of the entanglement spectrum fixes the density matrix up to unitarity transformation, we can say that it fixes the state of the block.   It is also clear that the Renyi entropy and the entanglement spectrum are also very closely related. In fact, if we know $\zeta$-function of $\rho_A$:
\be
\zeta_{\rho_A}(\alpha) \equiv \mbox{tr} \rho_{A}^{\alpha} =  \sum_{n=0}^{\infty} g_n \lambda_n^{\alpha} \; .
  \label{zeta}
\ee
at all $\alpha$ then we can find the eigenvalues $\lambda_n$, which have the meaning of probabilities ($0< \lambda_n < 1$) and their multiplicities $g_n$. From the definition of the Renyi entropy (\ref{Salphadef}) we see that
\be
  \zeta_{\rho_A}(\alpha) = \eu^{(1-\alpha) S_R (\rho_{A}, \alpha)} \; .
  \label{mS}
\ee
The aim of this paper is to use this relation to calculate the spectrum of $\rho_A$ for the  anisotropic $XY$ model, for which analytical expressions for the Renyi entropy are known explicitly \cite{jin, renyipaper}. The $XY$ model is one of the simplest integrable models (since quasiparticle excitations are essentially free fermions), while still having an interesting and non-trivial phase diagram.
Von Neumann entropy of the XY model was  first  determined rigorously using Toeplitz determinant representation and Riemann-Hilbert techniques in \cite{jpaits} and later  using thermodynamic arguments  in \cite{pes}, and it was studied in \cite{ellipsespaper}. The method of \cite{jpaits} was generalized in \cite{imm} for more general quantum spins. The evaluation of the Renyi Entropy for the $XY$ model, which was done in \cite{jin, renyipaper},
uses again the Fisher-Hartwig formulae and the Riemann-Hilbert approach. Although this is the first time that these exact results are used to access the full spectrum of $\rho_A$, general behaviors have already been known, because of the underlying free fermionic structure.

In fact, in \cite{LiHaldane06}, in analogy with the formulae of standard statistical mechanics, it has been proposed to represent the reduced density matrix as $\rho_A =  \eu^{-\mathcal{H}_A}/ \mathcal{Z}$. For AKLT models of interacting spins this formula was rigorously proved and the entanglement spectrum was calculated for several of these systems (see \cite{hosho,ying}).
For essentially non-interacting models, such as the $XY$ spin chain, the entanglement spectrum is known to be equidistant and the degeneracy of the eigenvalues is  given by the number of ways a given energy can be realized by different excitations. This is essentially a partitioning problem and it has been addressed already several years ago in an effort toward the optimization of Density Matrix Renormalization Group approaches \cite{DMRG-1999}. Our exact approach will agree with the results of \cite{DMRG-1999}.

It should also be mentioned that for some spin models the reduced density matrix was evaluated explicitly.

\section{Quantum entropies for the $XY$ model}

\begin{figure}
 \includegraphics{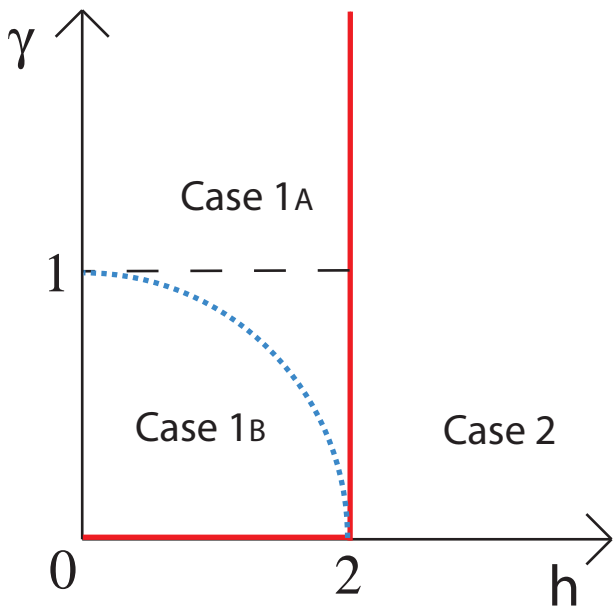}
  \caption {Phase diagram of the anisotropic $XY$ model in a constant magnetic field (only $\gamma \ge 0$ and $h \ge 0$ shown). The three cases {\sc $2$,
$1$a, $1$b}, considered in this paper, are clearly marked. The critical phases ($\gamma = 0$, $h \le 2$ and $h = 2$) are drawn in bold lines (red, online).
The boundary between cases {\sc $1$a} and {\sc $1$b}, where the ground state is given by two degenerate product states, is shown as a dotted line
(blue, online). The Ising case ($\gamma = 1$) is also indicated, as a dashed line.}
 \label{phasediagram}
\end{figure}

The anisotropic $XY$ spin chain is defined by the Hamiltonian
\be
   H=-\sum_{j=-\infty}^{\infty} \left[
   (1+\gamma)\sigma^x_{j}\sigma^x_{j+1}+(1-\gamma)\sigma^y_{j}\sigma^y_{j+1}
   + h\sigma^z_{j} \right] \; ,
   \label{xxh}
\ee
where $\gamma$ is the anisotropy parameter, $\sigma^x_j$, $\sigma^y_j$
and $\sigma^z_j$ are the Pauli matrices and $h$ is the magnetic field. The Hamiltonian is clearly symmetric under the transformations $\gamma \to - \gamma$ or $h \to -h$, therefore we can consider just the quadrant $\gamma >0$ and $h \ge0$ without loss of generality.
The system is gapped in the bulk of the phase diagram and has two phase transitions where the spectrum becomes critical: at $\gamma =0, |h|<2$ one has the $XX$ model (universality of free fermions on a lattice) and $|h|=2$ is the critical magnetic field of the Ising phase transition, see Figure \ref{phasediagram}.

The model was solved in \cite{Lieb,mccoy,mccoy2,gallavotti}. It is known that its correlation functions can be calculated using methods of Toeplitz determinants and Riemann-Hilbert problems. These techniques were applied in \cite{jpaits, jin, ellipsespaper, renyipaper} to the study of the quantum entropies.

The density matrix of the unique ground state $| GS \rangle$ of the model is
given by \mbox{$\rho_{AB}=|GS\rangle \langle GS|$}. The reduced density matrix
of a subsystem A is \mbox{$\rho_A= Tr_B(\rho_{AB})$}. We take the subsystem $A$ to be a block of $n$ consecutive spins (system $B$ is the state of the rest of the chain) and consider the double scaling limit $1<<n<<N=\infty$, where $N$ is the total number of sites in the chain, which we take to be infinite. Matrix elements of $\rho_{A}$ are correlation functions, see formulae (17) and (18) of \cite{jin}.

The R\'enyi entropy (\ref{Salphadef}) converges to the von Neumann entropy (\ref{Sdef}) for $\alpha \to 1$, therefore we can concentrate just on the former quantity. Its analytical expressions were derived in \cite{renyipaper} and they can be written as
\be
   S_R(\rho_{A},\alpha) = {1 \over 6} \; { \alpha \over 1 - \alpha} \; \ln \left( k \; k' \right)
   + {1 \over 3} \; { 1 \over 1 - \alpha} \; \ln \left(
   \theta_3^2 (0| \alpha \ii \tau_{0}) \over
   { \theta_2 (0| \alpha \ii \tau_{0}) \; \theta_4 (0| \alpha i \tau_{0}) }
   \right) \; + {1 \over 3} \ln 2 \; ,
   \label{renfinal2new}
\ee
for $h> 2$
and
\be
   S_R(\rho_{A},\alpha) = {1 \over 6} \; {\alpha \over 1 - \alpha } \; \ln
   \left( {k'\over k^2 } \right) + {1 \over 3} \; {1 \over 1 - \alpha} \;
   \ln \left( { \theta_2^2 (0| \alpha i\tau_{0})
   \over \theta_3 (0| \alpha i \tau_{0}) \; \theta_4 (0| \alpha i \tau_{0})  } \right) \;  + {1 \over 3} \ln 2 \; ,
   \label{renfinal4new}
\ee
for $h < 2$; where
\be
   \tau_0 \equiv \frac{I(k')}{I(k)} \; ,
   \qquad \qquad k'=\sqrt{1-k^2} \; ,
   \label{taudef}
\ee
$I(k)$ is the complete elliptic integral of the first kind,
\be
   I(k) = \int_{0}^{1}\frac{dx}{\sqrt{(1-x^2)(1 - k^{2}x^{2})}}
   \label{ellint}
\ee
and
\be
   \theta_{j}(z|\tau):= \theta_{j}(z,q) \; ,\quad q =  \eu^{\pi \ii \tau}
   \quad j = 1, 2, 3, 4,
   \label{thetatau}
\ee
are the elliptic theta functions defined by the following
Fourier series ($|q| < 1$)
\bea
   \theta_{1} (z, q) & = &  \ii \sum_{m=-\infty}^{\infty}
   (-1)^m q^{\left(\frac{2m-1}{2}\right)^2}
   \eu^{2 \ii z \left(m-\frac{1}{2}\right)} \; ,
   \label{t1} \\
   \theta_{2} (z, q) & = & \sum_{m=-\infty}^{\infty}
   q^{\left(\frac{2m-1}{2}\right)^2}
   \eu^{2 \ii z \left(m-\frac{1}{2}\right)} \; ,
   \label{t2} \\
   \theta_{3} (z, q) & = & \sum_{m=-\infty}^{\infty}
   q^{n^2} \eu^{2 \ii z m} \; ,
   \label{t3} \\
   \theta_{4} (z, q) & = & \sum_{m=-\infty}^{\infty}
   (-1)^m q^{m^2} \eu^{2 \ii z m} \;.
   \label{t4}
\eea
The elliptic parameter $k=k(\gamma,h)$ is defined in the different regions of the phase diagram as
\bea
  \qquad \qquad k \equiv \left\{ \begin{array}{ll}
  \sqrt{(h/2)^2+\gamma^2-1}\; /\; \gamma \; ,&
  \mbox{Case 1a:~$4(1-\gamma^2)<h^2<4$;} \\
  \sqrt{({1-h^2/4-\gamma^2})/({1-h^2/4})} \; , &
  \mbox{Case 1b:~$h^2<4(1-\gamma^2)$;} \\
  \gamma\; / \;\sqrt{(h/2)^2+\gamma^2-1} \; , &
  \mbox{Case 2~:~$h>2$.}
  \end{array} \right.
  \label{kmain}
\eea

Alternatively, we can write the R\'enyi in terms of the {\it $\lambda$ -
modular function} (see \cite{renyipaper}) as
\be
   S_R = \left\{ \begin{array}{lr}
      \displaystyle{
      {1 \over 6} \; { \alpha \over 1- \alpha } \; \ln \left( k \; k' \right)
      - {1 \over 12} \; { 1 \over 1-\alpha} \; \ln \Bigl[
      \lambda(\alpha \ii \tau_{0})
      \left( 1-\lambda(\alpha \ii \tau_{0}) \right) \Bigr]
      + {1 \over 3} \ln 2 }
      & h> 2 \cr
      \displaystyle{
      {1 \over 6} \; {\alpha \over 1-\alpha } \; \ln \left( {k'\over k^2 } \right)
      - {1 \over 12} \; {1 \over 1-\alpha } \;
      \ln \left[ \frac{1-\lambda(\alpha \ii \tau_{0})}{\lambda^2(\alpha \ii
      \tau_{0})} \right]
      + {1 \over 3} \ln 2 }
      & h<2 \cr
   \end{array} \right. \; .
   \label{renlambda}
\ee
The modular function is defined as ($\Im \tau > 0$)
\be
  \lambda(\tau) =  \frac{\theta^{4}_{2}(0|\tau)}{\theta^{4}_{3}(0|\tau)}
  \equiv k^2( \tau) \; , \quad \mbox{and} \quad
  1-\lambda(\tau) = \frac{\theta^{4}_{4}(0|\tau)}{\theta^{4}_{3}(0|\tau)}
  \equiv {k'}^2(\tau) \; ,
  \label{kappadef}
\ee
We note that the basic modular parameter $k \equiv k(\gamma, h)$ defined in (\ref{kmain}) coincide with the value of the function $k(\tau)$ at $ \tau = \ii \tau_0$.

A third representation of the Renyi entropy in terms of $q-$series will be useful to determine the multiplicities of the reduced density matrix eigenvalues \cite{renyipaper}:
\be
   S_R (\rho_A, \alpha) = \left\{ \begin{array}{lr}
      \displaystyle{{1 \over 12} { \alpha \over 1 - \alpha } \;
      \ln \left( {k^2 k'^2 \over 16 q } \right)
      + {2 \over 1 - \alpha } \ln \prod_{m=0}^\infty
      \left[ 1 + q^{(2m+1) \alpha} \right]}
      & h>2 \cr
      \displaystyle{{1 \over 6} {\alpha \over 1 - \alpha} \;
      \ln \left( { 16 q k' \over k^2 } \right)
      + {2 \over 1 - \alpha } \ln \prod_{m=1}^\infty
      \left[ 1 + q^{2 m \; \alpha} \right]
      + \ln 2}
      & h<2 \cr
   \end{array} \right. \; ,
   \label{renyiq}
\ee
where
\be
   q \equiv \eu^{- \pi \tau_0} = \eu^{- \pi I(k')/I(k)} \; .
\ee

\section{Spectrum of  $\rho_A$}

Using the expressions for the Renyi entropy we just listed, we now want to determine the eigenvalues $\lambda_n$ ($0<\lambda_n<1$) of the operator $\rho_A$ and their multiplicities $g_n$, through its momentum function (\ref{zeta}) using (\ref{mS}). Using (\ref{renyiq}) we have
\be
   \zeta_{\rho_A}(\alpha) = \left\{ \begin{array}{lr}
      \displaystyle{\eu^{\alpha\left(\frac{\pi \tau_0}{12}
      +\frac{1}{6}\ln\frac{k\;k'}{4}\right)}
      \prod_{m=0}^{\infty}\left(1 + q_{\alpha}^{2m+1}\right)^2}
      & h > 2 \cr
      \displaystyle{2 \eu^{\alpha\left(-\frac{\pi \tau_0}{6} +\frac{1}{6}\ln\frac{k'}{4k^2}\right)}
      \prod_{m=1}^{\infty} \left(1 + q_{\alpha}^{2m}\right)^2}
      & h<2 \cr
   \end{array} \right. \; ,
   \label{zeta1}
\ee
where
\be
   q_{\alpha} \equiv \eu ^{-\alpha\pi\tau_0} = q^\alpha \; .
   \label{qalpha}
\ee
To use these expression, we will need some results on q-series and elementary notions of the theory of partitions.

Let us concentrate first on the case $h>2$.  Classical arguments of the theory of partitions (see e.g. \cite{andrews}) tell us that
\be
   \label{fourier1}
   \prod_{n=0}^{\infty}\left(1 + q^{2n+1}\right) =
   \sum_{n=0}^{\infty}p_{\mathcal{O}}^{(1)}(n)q^n,
\ee
where $p_{\mathcal{O}}^{(1)}(0) = 1$ and $p_{\mathcal{O}}^{(1)}(n)$, for $n >1$, denotes the number of partitions of  $n$ into distinct positive {\it odd} integers, i.e.
\bea
   p_{\mathcal{O}}^{(1)}(n) & \equiv &
   \#\big\{(m_1, \ldots, m_k): m_j = 2r_j + 1, \quad m_1 > m_2 > \ldots > m_k,
   \\
   && \qquad n = m_1 + m_2 +  \ldots + m_k \big\} \; .
   \nonumber
\eea
Hence (\ref{zeta1}) for $h>2$ becomes
\be
   \zeta_{\rho_A}(\alpha) = \eu^{\alpha\left(\frac{\pi \tau_0}{12} +\frac{1}{6}\ln\frac{k\;k'}{4}\right)}
   \sum_{n=0}^{\infty} a_n q_{\alpha}^{n} \; ,
   \label{zeta2}
\ee
where,
\be
a_0 = 1 \; , \qquad
a_n = \sum_{l=0}^{n}p_{\mathcal{O}}^{(1)}(l)p_{\mathcal{O}}^{(1)}(n-l)
 \label{adef}
\ee

Since
\be
   q_{\alpha}^n = \left(\eu^{-\pi\tau_0n}\right)^{\alpha},
   \label{lambdan}
\ee
we conclude that
\be
   \zeta_{\rho_A}(\alpha)  = \sum_{n=0}^{\infty} a_n \lambda_n^{\alpha}, \qquad \lambda_n =
   \eu^{-\pi\tau_0n +\frac{\pi \tau_0}{12} + \frac{1}{6}\ln \frac{k\;k'}{4}}.
   \label{zeta3}
\ee
Comparing the last equation with (\ref{zeta}) we arrive at the following theorem:
\begin{theorem}
   Let the magnetic field $h >2$. Then, the eigenvalues of the reduced density matrix $\rho_{A}$ are given by
   \be
      \lambda_n = \eu^{\frac{1}{6}\ln \frac{k\;k'}{4} - \pi {I(k') \over I(k)} \left[ n - \frac{1}{12} \right] } , \qquad n = 0, 1, 2, \ldots
      \label{spectrumfinal}
   \ee
   and the corresponding multiplicities $g_n= a_n$ are defined by (\ref{adef}).
   \label{theorem1}
\end{theorem}

The case $h < 2$ is treated in a very similar way. Instead of (\ref{fourier1}) we use another combinatorial
identity,
\be
   \prod_{n=1}^{\infty}\left(1 + q^{2n}\right) = \sum_{n=0}^{\infty}p_{\mathcal{N}}^{(1)}(n)q^{2n},
   \label{fourier3}
\ee
where $p_{\mathcal{N}}^{(1)}(0) = 1$ and $p_{\mathcal{N}}^{(1)}(n)$, for $n >1$, denotes the number of partitions of  $n$ into distinct positive integers, i.e.
\bea
  p_{\mathcal{N}}^{(1)}(n) & \equiv &
  \#\big\{ (m_1, \ldots , m_k): m_1 > m_2 > \ldots > m_k \geq 0, \\
  && \qquad n = m_1 + m_2 +  \ldots + m_k \big\}.
  \nonumber
\eea
It is worth noticing that (see e.g.  \cite{andrews})
\be
   p_{\mathcal{N}}^{(1)}(n) = p_{\mathcal{O}}(n) \; ,
\ee
where $p_{\mathcal{O}}(n)$ denotes the partitions of $n$ into positive {\it odd} integers:
\be
  p_{\mathcal{O}}(n) \equiv
  \#\big\{ (m_1, \ldots , m_k): m_j = 2r_j + 1,
  \quad n = m_1 + m_2 +  \ldots + m_k \big\}.
\ee
The analog of equation (\ref{zeta2}), with the help of (\ref{fourier3}),  now reads as
\bea
   \zeta_{\rho_A} (\alpha) & = & 2 \eu^{\alpha\left(-\frac{\pi \tau_0}{6} +\frac{1}{6}\ln\frac{k'}{4k^2}\right)}
   \sum_{n=0}^\infty b_n q_\alpha^{2n}
   \label{zeta7} \\
   & = & 2 \sum_{n=0}^{\infty}
   b_n \lambda^{\alpha}_{n} , \qquad \lambda_n =
   \eu^{-2\pi\tau_0n -\frac{\pi \tau_0}{6} +\frac{1}{6}\ln\frac{k'}{4k^2}} ,
   \nonumber
\eea
where
\be
   b_0 = 1 \; , \qquad
   b_n = \sum_{l=0}^{n}
   p_{\mathcal{N}}^{(1)}(l) p_{\mathcal{N}}^{(1)}(n-l) \; .
   \label{bdef}
\ee

Finally, comparing (\ref{zeta7}) with equation (\ref{zeta}) we arrive at the
analog of Theorem \ref{theorem1} for the case $h<2$:
\begin{theorem}
   Let the magnetic field $h < 2$. Then, the eigenvalues of the reduced density matrix $\rho_{A}$ are given by the equation,
   \be
      \lambda_n = \eu^{\frac{1}{6}\ln\frac{k'}{4k^2}
      -2 \pi {I(k') \over I(k)} \left[ n + \frac{1}{12} \right] } ,
      \qquad n = 0, 1, 2, \ldots
      \label{spectrumfinal2}
   \ee
   and the corresponding multiplicities $g_n = 2b_n$ where the integers $b_n$ are determined by (\ref{bdef}).
   \label{theorem2}
\end{theorem}

\section{Asymptotics of  $g_n$}

Consider first the case $h >2$. Following the usual methodology, we introduce the generating function
\be
   f(z) := \sum_{n=0}^{\infty} g_n z^n \; .
   \label{genfunk}
\ee
This function is holomorphic in the unit disc.
Indeed, we have from (\ref{zeta2}) that,
\be
 f(z) =   \eu^{-\alpha \left(
   {1 \over 6} \ln {k \; k' \over 4} + {\pi \tau_0 \over 12} \right) }\zeta_{\rho_A} (\alpha), \quad
    \alpha = -\frac{1}{\pi \tau_0}\ln z.
\ee
Statement of holomorphicity then follows from the first equation in (\ref{zeta1}).

The function $f(z)$ has a singularity at $z=1$.  In order to see this,
we deduce from (\ref{mS})  the representation for $f(z)$ in terms of the entropy $S_{R}(\rho_A,\alpha)$
\be
   f(z) = \eu^{(1-\alpha)S_{R}(\rho_A,\alpha) -\alpha \left(
   \frac{1}{6}\ln\frac{k\;k'}{4} + \frac{\pi \tau_0}{12} \right) },
   \qquad \alpha = -\frac{1}{\pi \tau_0}\ln z \; .
   \label{fS}
\ee
In \cite{renyipaper}, using the explicit formulae (\ref{renfinal2new})
and  (\ref{renlambda}), and the modular properties of the $\lambda$ - function,
\be \label{kappamod1}
\lambda\left(-\frac{1}{\tau}\right) = 1 - \lambda(\tau), \qquad \lambda(\tau + 1) = \frac{\lambda(\tau)}{\lambda(\tau) - 1},
\ee
 it was obtained that
\bea
   S_{R}(\rho_{A},\alpha) & = &
   \frac{1}{\alpha(1-\alpha)} \frac{\pi}{12}\frac{I(k)}{I(k')}
   + \frac{\alpha}{1-\alpha}\frac{1}{6}\ln\frac{k \: k'}{4}
   + \Ord \left( \eu^{-\frac{\pi}{\alpha\tau_{0}}}\right) \; ,
   \label{smallalpha111}
   \\
   && \qquad \qquad \qquad \qquad \qquad \qquad
   \alpha \to 0, \quad -\frac{\pi}{2} < \arg \alpha < \frac{\pi}{2} \; .
   \nonumber
\eea
Hence,
\be
   f(z) = \eu^{-\frac{\pi^2}{12\ln z} + {1 \over 12} \ln z
   + \Ord \left( \eu^{\pi^2/ \ln z} \right)} ,
   \qquad z \to 1, \quad |z| < 1 \;.
   \label{fatzero}
\ee

\begin{figure}
   \includegraphics[width=0.45\columnwidth]{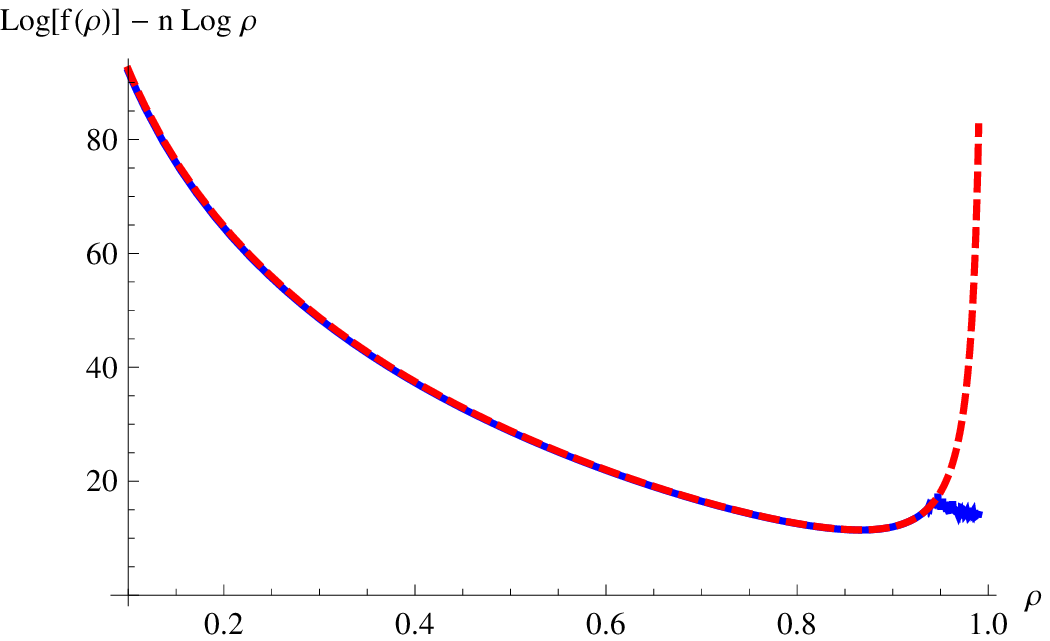}
   \hskip 0.5cm
   \includegraphics[width=0.45\columnwidth]{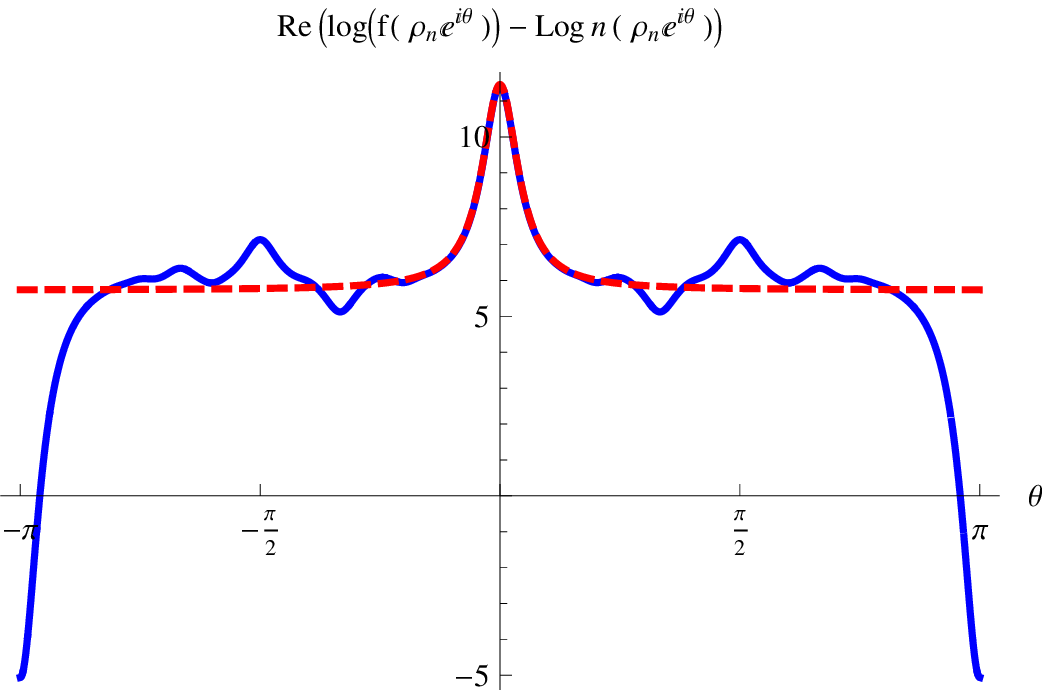}
   \caption {Case $h>2$: plots of the logarithm of the generating function in the Cauchy integral (\ref{Cauchy}) as a function of the radius $\rho$ (left panel) and of the angular phase $\theta$ at the saddle point radius $\rho_n$ given by (\ref{rhon1}) (right panel; only real part shown). The plots show the comparison between the exact expression (\ref{fS}) (continuous line, blue on-line) and its asymptotic approximation (\ref{fatzero}) (dashed line, red on-line) at $n=40$.}
   \label{plot1}
\end{figure}

The coefficients $g_n$ are given by the Cauchy formula,
\be
   g_n = \frac{1}{2\pi \ii} \int_{|z|= 1-\epsilon} \frac{f(z)}{z^{n+1}} \,
   \de z \; ,
   \label{Cauchy}
\ee
From this formula, the large $n$ asymptotics of $g_n$ can be rigorously obtained by the Hardy-Ramanujan-Rademacher circle method (see e.g. \cite{rademacher}) using the modular properties (\ref{kappamod1}) of the $\lambda$-function.
According to the circle method, the leading contribution to  integral
(\ref{Cauchy}) comes from the neighborhood{\footnote{The implementation of the circle method in its full power
would yield  the  Hardy-Ramanujan-Rademacher type expansions for the multiplicities $g_n$ (cf.
\cite{rademacher} where the classical case of $p_{ \mathcal{N}}(n)$ is considered).
In this paper, we are only concerned with the leading
behavior of $g_n$, and to this end we only need the localization of the integral near the
point $z =1-\epsilon$. The rigorous proof of this property of   integral (\ref{Cauchy})
is not at all trivial. Indeed, it needs again  the
modular properties of the function $f(z)$ .

It should be also mentioned that  there are more general techniques of the asymptotic analysis of the
partitions, such as Meinardus theorem (see e.g. \cite{andrews}; see also \cite{berndt}).
These techniques do not exploit the modular properties of the  corresponding  generating functions,
however, unlike the circle method,  they only provide the leading terms of the asymptotics.
The Meinardus theorem, as it is stated in \cite{andrews}, is not directly applicable to
generating function (\ref{genfunk}). }}
of the point $1 -\epsilon$.
This fact can be also demonstrated by plotting the function $g(z) = \ln f(z) - n \ln z$,
see Figure \ref{plot1}. Therefore, we can replace the explicit formula (\ref{Cauchy})
by the estimate,

\be
   g_{n} \simeq
   \frac{1}{2\pi \ii} \int_{\mathcal{L}} \frac{f(z)}{z^{n+1}} \, \de z
   \simeq \frac{1}{2\pi \ii} \int_{\mathcal{L}} {\eu^{-G(z)} \over z} \, \de z \;,
   \quad n \to \infty,
   \label{asymp1}
\ee
where ${\mathcal{L}} = \left\{|z| = \rho_n , \,\, |\arg z| < \delta \le - \ln \rho_n \right\}$, and
\be
   G(z) \equiv \frac{\pi^2}{12\ln z} + \left( n - {1 \over 12} \right) \ln z \; .
\ee
We  determine $\rho_n$ as the stationary point of $G(z)$, i.e.:
\be
   \left. {d G(z) \over dz} \right|_{z = \rho_n} = 0
   \qquad \Rightarrow \qquad
   \rho_n = \eu^{- {\pi \over \sqrt{12 n - 1}}} \equiv \eu^{-\epsilon_n} \; .
   \label{rhon1}
\ee
Switching then to polar coordinate $z = \eu^{-\epsilon_n + \ii \theta}$, we can rewrite
$G(z)$ in the form,
$$
G(z) = \frac{\pi^2}{12}(-\epsilon_n + \ii \theta)^{-1} + \left(n - \frac{1}{12}\right)(-\epsilon_n + \ii \theta)
$$
\be\label{anasymp2}
= -\frac{\pi}{\sqrt{3}}\sqrt{n - \frac{1}{12}} + \frac{\pi^2}{12\epsilon^3_n}\theta^2\left(1 +\Ord\left(\frac{\theta}{\epsilon_n}\right)\right),
\quad z \in \mathcal{L}.
\ee
It can be shown, using again  the circle method, that there exists a positive $\kappa_0$ such that
the following choice of the parameter $\delta$ in the definition of the arc $\mathcal{L}$ is consistence with
estimate  (\ref{asymp1}):
\be\label{deltakappa}
\delta = \epsilon_n^{1 +2\kappa}\;, \quad  0< \kappa < \kappa_0.
\ee
Using this specification of $\delta$ we  re-write  estimate (\ref{anasymp2})
as
$$
G(z)= -\frac{\pi}{\sqrt{3}}\sqrt{n - \frac{1}{12}} + \frac{\pi^2}{12\epsilon^3_n}\theta^2\Bigl(1 +\Ord\left(\epsilon_n^{2\kappa}\right)\Bigr),
$$

\be
= -\frac{\pi}{\sqrt{3}}\sqrt{n - \frac{1}{12}} + \frac{\pi^2}{12\epsilon^3_n}\theta^2\Bigl(1 +\Ord\left(n^{-\kappa}\right)\Bigr)\;,\quad
0< \kappa < \kappa_0\;, \quad n \to \infty, \quad z \in \mathcal{L}.
\ee
In its turn, this estimate yields the following representation for $G(z)$ on  $\mathcal{L}$ ,
\be\label{Gnew}
G(z)= -\frac{\pi}{\sqrt{3}}\sqrt{n - \frac{1}{12}} + \frac{\pi^2}{12\epsilon^3_n}t^2(\theta),
\ee
where $ t(\theta)$ is a function holomorphic in the neighborhood of the interval $[-\delta, \delta]$ and  satisfying the estimates,

\be\label{test}
t(\theta) = \theta \Bigl(1 +\Ord\left(n^{-\kappa}\right)\Bigr),\quad \frac{dt}{d\theta} = 1 +\Ord\left(n^{-\kappa}\right),
\ee

$$
0< \kappa < \kappa_0\;, \quad n \to \infty, \quad \theta \in [-\delta, \delta].
$$

From (\ref{Gnew}) we have that
\be\label{last01}
\frac{1}{2\pi \ii} \int_{\mathcal{L}} {\eu^{-G(z)} \over z} \, \de z
=\eu^{{\pi \over \sqrt{3}} \sqrt{ n -{1 \over 12} }}{1\over 2 \pi} \int_{-\delta}^{\delta}
  \eu^{- {\pi^2 \over 12 \epsilon_n^3}t^2(\theta) } \de \theta .
\ee
At the same time, equations (\ref{test}) allow us to  use $t=t(\theta)$ as a new integration variable and transform
(\ref{last01}) into the asymptotic formula,
$$
\frac{1}{2\pi \ii} \int_{\mathcal{L}} {\eu^{-G(z)} \over z} \, \de z
= \eu^{{\pi \over \sqrt{3}} \sqrt{ n -{1 \over 12} }}{1\over 2 \pi} \int_{t(-\delta)}^{t(\delta)}
  \eu^{- {\pi^2 \over 12 \epsilon_n^3}t^2 }\frac{d\theta}{dt} \de t
$$

$$
= \eu^{{\pi \over \sqrt{3}} \sqrt{ n -{1 \over 12} }}\left[{1\over 2 \pi} \int_{-\delta}^{\delta}
  \eu^{- {\pi^2 \over 12 \epsilon_n^3}t^2 }\frac{d\theta}{dt} \de t + \Ord\left(n^{-\frac{1+4\kappa}{2}}
  \eu^{-c_{\kappa}n^{(1-4\kappa)/2}}\right)\right]
  \quad \left(c_{\kappa}
  = \left(\frac{\pi^2}{12}\right)^{\frac{1+4\kappa}{2}}
> 0\right)
$$

\be\label{last1}
=\eu^{{\pi \over \sqrt{3}} \sqrt{ n -{1 \over 12} }}{1\over 2 \pi}\left[ \int_{-\delta}^{\delta}
  \eu^{- {\pi^2 \over 12 \epsilon_n^3}t^2 } \de t \Bigl(1 +\Ord\left(n^{-\kappa}\right)\Bigr) + \Ord\left(n^{-1}\right)
  \right], \quad n \to \infty,
\ee

Assume now that {\footnote{It is worth noticing, that under condition (\ref{kappacond}) the term $\Ord\left(n^{-1}\right)$
in (\ref{last1}) becomes in fact $\Ord\left(\eu^{-c_{\kappa}n^{\beta}}\right)$, $\beta >0$ , that is $\Ord\left(n^{-\infty}\right)$.}}
\be\label{kappacond}
0 < \kappa < \min\left\{\frac{1}{4}, \kappa_0\right\}.
\ee
Then, the integral in the right hand side of (\ref{last1}) can be estimated as follows,
$$
{1\over 2 \pi} \int_{-\delta}^{\delta}
  \eu^{- {\pi^2 \over 12 \epsilon_n^3} t^2 } \de t
= {1\over 2 \pi} \int_{-\infty}^{\infty}
  \eu^{- {\pi^2 \over 12 \epsilon_n^3} t^2 } \de t
  - {1\over  \pi} \int_{\delta}^{\infty}
  \eu^{- {\pi^2 \over 12 \epsilon_n^3} t^2 } \de t
$$

$$
= {1\over 2 \pi} \int_{-\infty}^{\infty}
  \eu^{- {\pi^2 \over 12 \epsilon_n^3} t^2 } \de t
  - \epsilon_n^{3/2}{1\over  \pi} \int_{\delta\epsilon_n^{-3/2}}^{\infty}
  \eu^{- {\pi^2 \over 12 }t^2 } \de t
  = {1\over 2 \pi} \int_{-\infty}^{\infty}
  \eu^{- {\pi^2 \over 12 \epsilon_n^3} t^2 } \de t +\Ord\left(n^{-\infty}\right)
$$

\be\label{last2}
= {1\over 2 \pi}\sqrt{{12\epsilon_n^{3}\over\pi}}+\Ord\left(n^{-\infty}\right) =
2^{-3/2} 3^{-1/4} \left( n - {1 \over 12} \right)^{-3/4} + \Ord\left(n^{-\infty}\right)
\ee
Estimates (\ref{asymp1}), (\ref{last1}), and (\ref{last2}) yield the following asymptotic
formula for the multiplicities $g_n$ of the eigenvalues of the reduced density matrix for $h >2$.

 \be
     g_n \simeq 2^{-3/2} 3^{-1/4} n^{-3/4} \eu^{\pi \sqrt{\frac{n}{3}}} \; , \qquad n \to \infty \; .
     \label{asymp12}
  \ee

Turning now to the $h<2$ case, using (\ref{zeta7}) and remembering that $g_n = 2 b_n$, we have
\be
   \zeta_{\rho_A} (\alpha) = \eu^{\alpha \left(
   {1 \over 6} \ln {k'\over 4 k^2} - {\pi \tau_0 \over 6} \right) }
   f \left( \eu^{- 2\pi \tau_0 \alpha} \right) ,
\ee
with $f(z)$ defined as in (\ref{genfunk}). As before, using (\ref{mS}) we can express $f(z)$ in terms of the Renyi Entropy $S_R (\rho_A, \alpha)$:
\be
   f(z) = \eu^{(1-\alpha)S_{R}(\rho_A,\alpha) -\alpha \left(
   \frac{1}{6}\ln\frac{k'}{4 k^2} - \frac{\pi \tau_0}{6} \right) },
   \qquad \alpha = -\frac{1}{2 \pi \tau_0}\ln z \; .
   \label{fS1}
\ee

\begin{figure}
   \includegraphics[width=0.45\columnwidth]{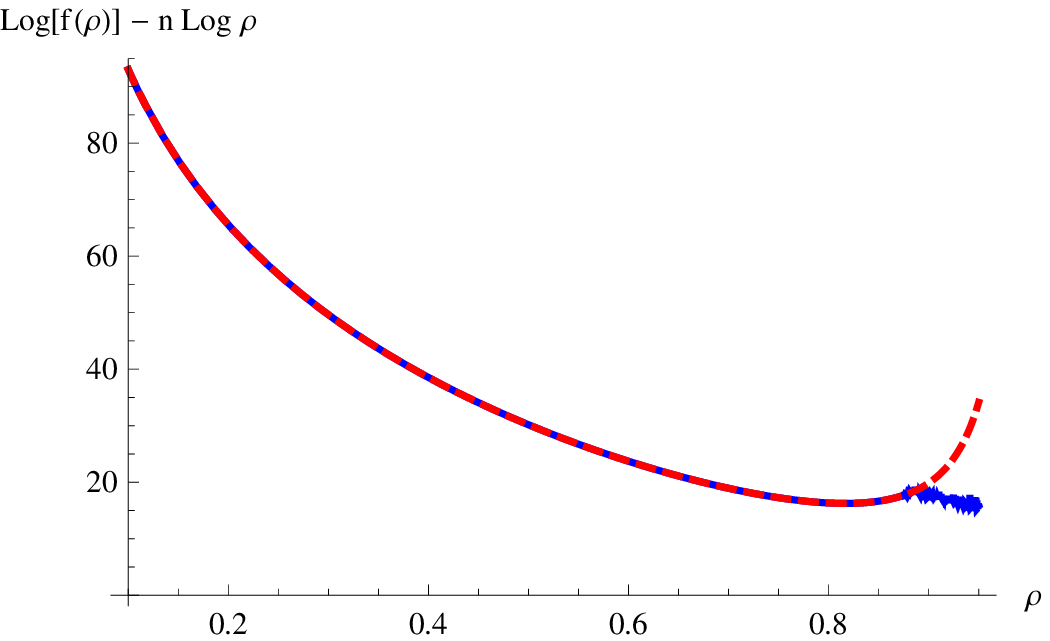}
   \hskip 0.5cm
   \includegraphics[width=0.45\columnwidth]{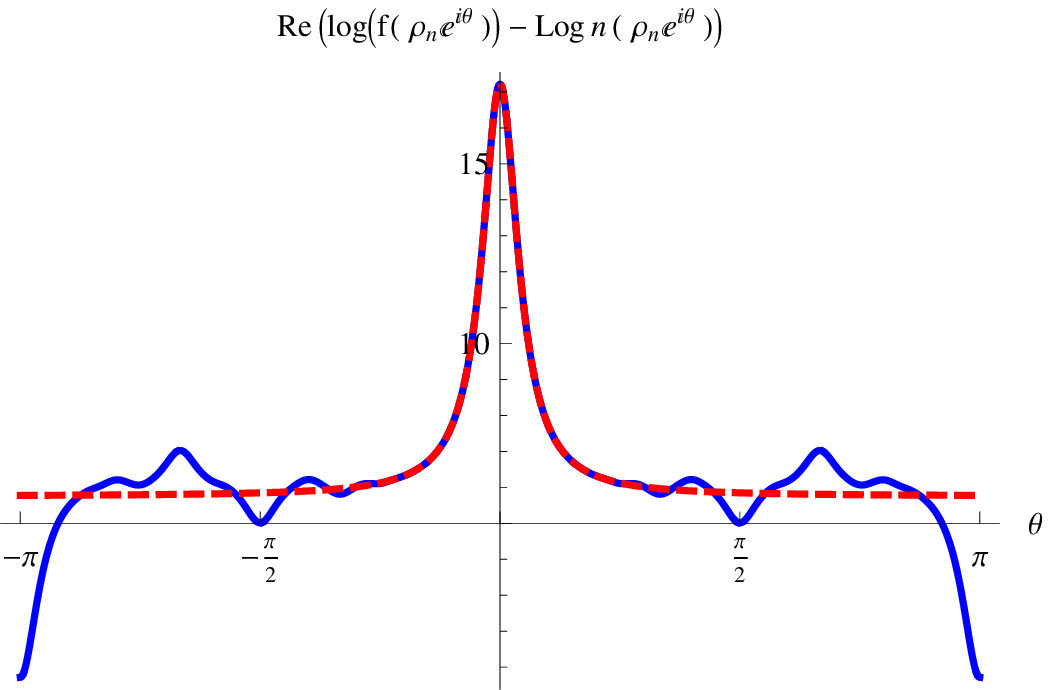}
   \caption {Case $h<2$: plots of the logarithm of the generating function in the Cauchy integral (\ref{Cauchy}) as a function of the radius $\rho$ (left panel) and of the angular phase $\theta$ at the saddle point radius $\rho_n$ given by (\ref{rhon2}) (right panel; only real part shown). The plots show the comparison between the exact expression (\ref{fS1}) (continuous line, blue on-line) and its asymptotic approximation (\ref{fatzero1}) (dashed line, red on-line) at $n=40$.}
   \label{plot2}
\end{figure}

Again, we are interested in the neighbors of the point $z \sim 1$ (see Figure \ref{plot2}), where we can use the asymptotics derived in \cite{renyipaper}:
\bea
   S_{R}(\rho_{A},\alpha) & = &
   \frac{1}{\alpha(1-\alpha)} \frac{\pi}{12}\frac{I(k)}{I(k')}
   + \frac{\alpha}{1-\alpha}\frac{1}{6}\ln \frac{k'}{4k^2}
   + \Ord \left( \eu^{-\frac{\pi}{\alpha\tau_{0}}}\right) \; ,
   \label{smallalpha2}
   \\
   && \qquad \qquad \qquad \qquad \qquad \quad
   \alpha \to 0, \quad -\frac{\pi}{2} < \arg \alpha < \frac{\pi}{2} \; .
   \nonumber
\eea
Using (\ref{smallalpha2}) in (\ref{fS1}) we have (cf. (\ref{fatzero}))
\be
   f(z) = \eu^{-\frac{\pi^2}{6\ln z} - {1 \over 12} \ln z
   + \Ord \left( \eu^{2\pi^2/ \ln z} \right)} ,
   \qquad z \to 1, \quad |z| < 1 \; .
   \label{fatzero1}
\ee

The calculation proceeds exactly as before,  where in   (\ref{asymp1}) we now have
\be
   G(z) \equiv \frac{\pi^2}{6\ln z} + \left( n + {1 \over 12} \right) \ln z \; ,
\ee
and the saddle point $z = \rho_n$ is:
\be
   \left. {dG(z) \over d z} \right|_{z = \rho_n} = 0
   \qquad \Rightarrow \qquad
   \rho_n = \eu^{- {\pi \sqrt{2} \over \sqrt{12 n + 1}}} \equiv \eu^{-\epsilon_n} \; .
   \label{rhon2}
\ee
Instead of (\ref{last1}) we now have,
\be\label{last12}
\frac{1}{2\pi \ii} \int_{\mathcal{L}} {\eu^{-G(z)} \over z} \, \de z
=\eu^{{\pi  \sqrt{\frac{2}{3}}\left( n +{1 \over 12}\right) }}\left[{1\over 2\pi} \int_{-\delta}^{\delta}
  \eu^{- {\pi^2 \over 6\epsilon_n^3} t^2 } \de t \Bigl(1 + \Ord\left(n^{-\kappa}\right)\Bigr)
  +\Ord\left(n^{-1}\right)\right],
\ee
with $\delta$ defined by exactly the same equation (\ref{deltakappa}). Choosing $\kappa$ as in (\ref{kappacond}), we can again approximate the integral in the right hand side of (\ref{last12}) by a complete Gaussian integral (cf. \ref{last2}):

$$
{1\over 2 \pi} \int_{-\delta}^{\delta} \eu^{- {\pi^2 \over 6 \epsilon_n^3} t^2 } \de t
  = {1\over 2\pi} \int_{-\infty}^{\infty}
  \eu^{- {\pi^2 \over 6 \epsilon_n^3} t^2 } \de t +\Ord\left(n^{-\infty}\right)
$$

\be\label{last22}
= {1\over 2 \pi}\sqrt{{6\epsilon_n^{3}\over\pi}}+\Ord\left(n^{-\infty}\right) =
2^{-5/4} 3^{-1/4} \left( n + {1 \over 12} \right)^{-3/4} + \Ord\left(n^{-\infty}\right)
\ee
Estimates (\ref{asymp1}), (\ref{last12}), and (\ref{last22}) yield the following asymptotic
formula\footnote{It is worth noticing that for the case
$h < 2$ the generating function $f(z)$ can be easily transformed to the one satisfying the conditions of the Menardus theorem and hence asymptotics (\ref{asymp23}) can be also obtained by using the Menardus theorem.} for the multiplicities $g_n$ of the eigenvalues of the reduced density matrix for $h <2$.
\be
     g_n \simeq 2^{-5/4} 3^{-1/4} n^{-3/4} \eu^{\pi \sqrt{\frac{2}{3} \; n}} \; , \qquad n \to \infty \; .
     \label{asymp23}
\ee

\section{Critical lines}

As we mentioned in the introduction, there are two critical lines in the phase diagram of the $XY$ model, where the gap closes: at the critical magnetic field $h=2$ and at the isotropic line $\gamma = 0$ \footnote{Also known as the $XX$ spin chain.} and $|h|<2$.
In critical cases, the  $\zeta$ function behaves as
\be
   \zeta_{\rho_A} (\alpha) = \Gamma_\alpha \;
   \xi^{-{c \over 6} \left( \alpha - 1/\alpha \right)} \; ,
   \label{pasqzeta}
\ee
where $\Gamma_\alpha$ is a non-universal constant and $\xi$ is the relevant length-scale in the considered regime.
This was first discovered for the isotropic case \cite{jin} \footnote{The Renyi entropy was first  calculated for $XX$ spin chain} and confirmed by conformal field theory \cite{calabreselefevre}.
The straightforward application of the asymptotic formulae found in \cite{renyipaper}, shows agreement with (\ref{pasqzeta}) close to the critical lines, with the length-scale set by the inverse energy gap $\Delta$, i.e.
\bea
   \xi^{-1} = \Delta = |h-2| \; , \qquad & c={1\over2} \; , & \qquad h \to 2 \; , \\
   \xi^{-1} = \Delta = |\gamma| \; , \qquad & c= 1 \; , & \qquad \gamma \to 0 \; , |h|<2 \; .
\eea

Our Theorems \ref{theorem1} and \ref{theorem2} for the eigenvalues distributions in the different regimes remain valid arbitrarily close to the critical lines: while the eigenvalues $\lambda_n$ tend to collapse and vanish with the energy gap, namely
\be
   \lambda_n \simeq \Delta^{c \over 6} \; \eu^{{\pi^2 \over \ln \Delta} n}\; , \qquad \Delta \to 0 \; ,
\ee
their multiplicities do not depend on the parameters $\gamma$ and $h$ of the model. This agrees with \cite{moore}.

\section{Conclusions}

We have calculated the spectrum of  reduced density matrix of a large block of spins in the ground state of $XY$ spin chain [entanglement spectrum  \cite{LiHaldane06} ], using our  results on  Renyi entropy \cite{renyipaper}.

We have confirmed the expectation that, being the model essentially non-interacting, the eigenvalues are equidistant and their multiplicities have simple interpretation in terms of combinatorics and different partitions of integers, see Theorem \ref{theorem1} and \ref{theorem2}.
The exact formulae for the eigenvalues have been given in terms of the parameters of the model.

The asymptotic behavior of the multiplicities has been calculated, using the modular properties of the Renyi entropy.
The leading terms of the asymptotics are given in equations (\ref{asymp12}) and (\ref{asymp23}) for strong
and weak magnetic field, respectively.  Our results agree with the estimates in \cite{DMRG-1999}. However, this is not the log-normal behavior quoted in \cite{pescheleisler} for the $XY$ model and also taken from \cite{DMRG-1999}. In fact, the log-normal result was achieved by combining (and smearing) the degeneracy with the eigenvalue behavior to give an estimate of the behavior of an effective eigenvalue in a non-integrable system. This estimate is important to implement an efficient DMRG calculation for generic systems.

\section{Acknowledgments}
We would like to thank  S. Bravy,  B. McCoy  and P. Morton for discussions.
The project was supported in part  by NSF Grants  DMS 0905744, DMS-0705263 , and DMS-0701768 and by  PRIN Grant 2007JHLPEZ.

\appendix

\bibliographystyle{amsalpha}

\end{document}